\pdfoutput=1

\documentclass[11pt]{article}

\usepackage[]{acl}

\usepackage[algo2e]{algorithm2e} 
\usepackage{algorithm, algorithmic}

\usepackage[table]{xcolor}
\usepackage{tabularx}
\newcolumntype{C}{>{\centering\arraybackslash}X}
\usepackage{times}
\usepackage{arydshln}
\usepackage{ulem}
\usepackage{latexsym}
\usepackage{booktabs}
\usepackage{amsmath,amsfonts}
\usepackage{multirow}
\usepackage{arydshln}
\usepackage{colortbl}
\usepackage{booktabs}
\usepackage{tcolorbox}
\usepackage{multirow}
\usepackage{subcaption}
\usepackage[table,xcdraw]{xcolor}  
\usepackage{soul}
\usepackage{graphicx}
\definecolor{lightgreen}{RGB}{200,255,200}  
\newcommand{\hlgreen}[1]{\sethlcolor{lightgreen}\hl{#1}}
\newcommand{\hlpink}[1]{\sethlcolor{pink}\hl{#1}}
\definecolor{darkergreen}{rgb}{0.0, 0.5, 0.0}
\definecolor{darkgreen}{RGB}{0, 70, 0}
\definecolor{darkerred}{rgb}{0.5, 0.0, 0.0}
\usepackage[framemethod=TikZ]{mdframed}
\usepackage[T1]{fontenc}
\definecolor{revised}{RGB}{0, 0, 255}

\usepackage[utf8]{inputenc}

\usepackage{microtype}

\title{
\raisebox{-0.3\height}{\includegraphics[width=0.06\textwidth]{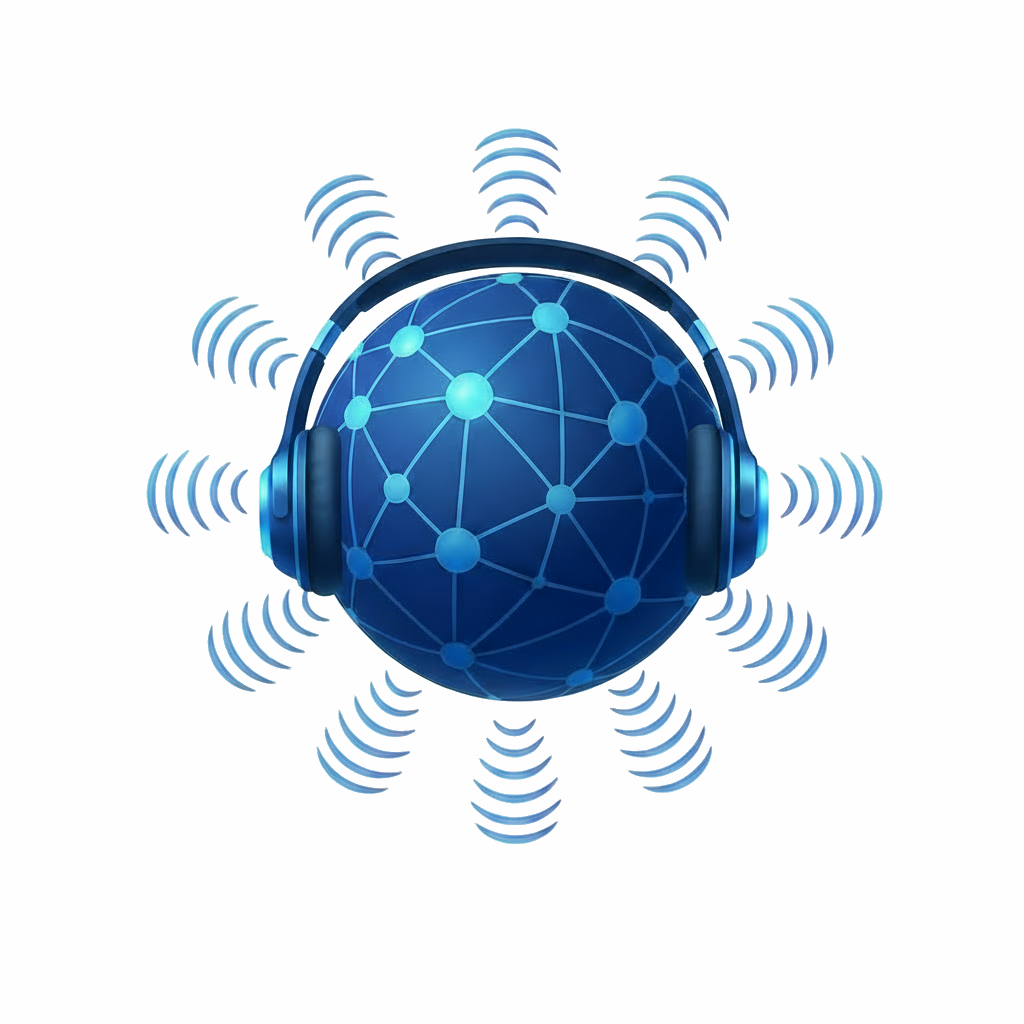}}%
\hspace{0.00em}
SPUR: A Plug-and-Play Framework for Integrating \underline{Sp}atial Audio \underline{U}nderstanding and \underline{R}easoning into Large Audio-Language Models}

\author{
    S Sakshi$^{12}$,
    Vaibhavi Lokegaonkar$^{1}$,
    Neil Zhang$^{2}$, \\
    \bf Ramani Duraiswami$^{1*}$,
    Sreyan Ghosh$^{1*}$,
    Dinesh Manocha$^{1*}$ ,
    Lie Lu$^{2*}$ \\
    University of Maryland, College Park, USA$^{1}$, Dolby Laboratories, USA$^{2}$ \\
    Correspondence: \{ssakshi, sreyang\}@umd.edu \\ \\
    Project: \url{https://sakshi113.github.io/spur/}
}

\begin{document}
\maketitle
\begin{abstract}

Spatial perception is central to auditory intelligence, enabling accurate understanding of real-world acoustic scenes and advancing human-level perception of the world around us. While recent large audio-language models (LALMs) show strong reasoning over complex audios, most operate on monaural inputs and lack the ability to capture spatial cues such as direction, elevation, and distance. We introduce {\textbf{SPUR}}, a lightweight, plug-in approach that equips LALMs with spatial perception through minimal architectural changes. SPUR consists of: (i) a First-Order Ambisonics (FOA) encoder that maps (W, X, Y, Z) channels to rotation-aware, listener-centric spatial features, integrated into target LALMs via a multimodal adapter; and (ii) {\textbf{SPUR-Set}}, a spatial QA dataset combining open-source FOA recordings with controlled simulations, emphasizing relative direction, elevation, distance, and overlap for supervised spatial reasoning. Fine-tuning our model on the SPUR-Set consistently improves spatial QA and multi-speaker attribution while preserving general audio understanding. SPUR provides a simple recipe that transforms monaural LALMs into spatially aware models. Extensive ablations validate the effectiveness of our approach.
\footnote{$^*$Equal advising.}\footnote{Work done by Sakshi during an internship at Dolby.}
\end{abstract}

\setlength{\abovedisplayskip}{6pt}
\setlength{\belowdisplayskip}{6pt}

\section{Introduction}
Understanding where sounds originate, how far they are, and how they move is central to machine listening in real environments—from multi-party interaction and telepresence to AR/VR and robot navigation. Spatial audio provides these cues explicitly, like interaural timing/level differences; HRTFs, and modern systems deploy a variety of spatial formats, notably scene-based First-Order Ambisonics (FOA), channel-based surround, and object-based renderers. Prior work spans two complementary thrusts: spatial understanding (detection, localization, separation, spatial scene consistency) and spatial generation (spatialization, ambisonics/binaural synthesis), together with the formats, datasets, and metrics that support them. Across these efforts, FOA consistently emerges as a compact, scene-aligned representation that captures global spatial structure, yet there remains a clear gap in methods that couple rich spatial audio perception with language-based reasoning. Contemporary large audio-language models (LALMs) are almost uniformly trained and evaluated on monaural inputs~\cite{gong2024listen,deshmukh2023pengi,qwen2audio}; they answer “what,” but routinely miss “where,” “how far,” and “in which relative arrangement,” limiting downstream abilities such as multi-speaker attribution, navigation, and audio-guided manipulation, etc.

Significant progress has been made in monaural audio comprehension with LALMs, including tasks like temporal reasoning, general AQA/captioning, etc. Some frontier LALMs include Audio Flamingo 2~\citep{ghosh2025audio}, Audio Flamingo 3~\citep{goel2025audio} GAMA~\citep{ghosh-etal-2024-gama}, SALMONN~\citep{tang2024salmonn}, Pengi, Qwen-Audio / Qwen-2-Audio and LTU by coupling audio encoders with decoder-only LLMs by gated cross-attention or prefixing strategies. Yet, despite the breadth of reasoning tasks, these models operate on monaural representations; explicit 3D spatial cues such as direction, elevation, and distance remain out of scope in both inputs and supervision, limiting downstream abilities like multi-speaker attribution and overlap-aware inference. 

Recent spatial LALMs fall into two major categories. 1) Binaural QA LALM: BAT~\citep{zheng2024bat} fuses a binaural spatial encoder (Spatial-AST) with LLaMA-2 and trains on synthetic binaural scenes; its SpatialSoundQA covers detection/DoA/distance and pairwise relations, but the answer space is yes/no or single-token answers over less than two static sources, which limits evaluation beyond accuracy and under-rewards multi-hop or relational reasoning. 2) FOA-centric QA LALM: Towards Spatial Audio Understanding via QA~\citep{sudarsanam2025spatialaudiounderstandingquestion} uses a real FOA dataset STARSS23~\citep{shimada2023starss23audiovisualdatasetspatial}, with scene-level QA derived from spatio-temporal captions, but only targets SELD-adjacent labels like presence, DoA bins, near/far, ordering, rather than open-ended multi-entity reasoning. Other approaches to infuse spatial perception into LLMs include injecting FOA intensity vectors into a Whisper→Q-Former→LLM pipeline and report strong self-supervised learning, primarily speech-only and not general QA ~\citep{tang2024largelanguagemodelsunderstand}. Other adjacent efforts like SALM~\citep{hu2025salm} introduce factorized spatial–semantic embeddings learned via contrastive pairing of FOA audio and text, enabling controllable direction editing and zero-shot DoA tagging in a shared language-aligned space. SING targets egocentric, on-device assistance, fusing real-time DoA cues with ASR/LLM modules under latency and power constraints to support speech-centric queries in the wild. These approaches are novel in representation control (SALM) and wearable deployment (SING~\citep{mishra2025spatialaudioprocessinglarge}), but neither provides a general spatial QA system over diverse, multi-event FOA scenes.
\vspace{1mm}

\noindent{\textbf{Our Contributions:}} In this work, we present \textbf{SPUR}, a lightweight plug-and-play approach that spatializes existing LALMs with minimal architectural changes while preserving their general audio understanding and reasoning capabilities. Our key idea is to treat FOA as a scene-aligned carrier of spatial cues and to introduce rotation-aware, listener-centric features into an LALM via a lightweight multimodal adapter on the audio encoder. To support training and evaluation, we further propose \textbf{SPUR-Set}, a task-focused spatial QA benchmark curated through a three-step pipeline with complex, reasoning-oriented QAs designed to teach essential spatial skills such as direction, elevation, distance, and overlap. Unlike prior work that primarily focuses on binaural input, our method generalizes to multichannel spatial audio. Extensive experiments and ablations show the efficacy of our approach. Our main contributions are:

\begin{enumerate}
    \item We propose \textbf{SPUR}, a parameter-efficient spatial adapter that endows LALMs with fine-grained spatial perception. SPUR consists of a lightweight FOA encoder that transforms first-order Ambisonics $(W, X, Y, Z)$ signals into rotation-aware, listener-centric representations, and an adapter that conditions the target LALM without retraining its core language components. This introduces plug-and-play spatial reasoning while preserving general audio understanding. Fine-tuning with our proposed dataset consistently enhances spatial QA and multi-speaker attribution, while maintaining or slightly improving non-spatial performance, thereby demonstrating that spatial conditioning acts as an additive capability rather than a trade-off.

    \item We introduce \textbf{SPUR-Set}, a hybrid dataset combining real FOA recordings and controlled simulations, designed to emphasize spatial reasoning and multi-speaker attribution across complex, multi-event acoustic scenes. The dataset includes spatially grounded audio-caption pairs and QAs across six expert reasoning skills, providing a robust foundation for training spatially aware ALMs.

    \item We conduct ablations on (a) FOA feature design and rotation handling, (b) adapter placement/width, and (c) supervision mixtures (real FOA vs. simulation), and provide a structured comparison against language-grounded supervision to clarify what drives spatial gains.
\end{enumerate}
\section{Related Work}
\label{sec:related}
\textbf{Large Audio Language Models (LALMs):} Large Audio-Language Models (LALMs) have seen rapid progress in perceptual and reasoning capabilities. Early works such as LTU, SALMONN, GAMA, Pengi, and Phi-4-MM~\citep{phi4} focused on short, monaural inputs and supported only a limited set of perception and reasoning tasks. More recent systems, including Qwen-2-Audio,and Audio Flamingo~\citep{kong2024audio}, expanded the skill set and extended input handling to longer audio sequences, with models like Audio Flamingo 2, Step-2-Audio~\citep{wu2025stepaudio2technicalreport}, and Kimi-Audio~\citep{kimi-audio} enabling robust processing of long-form inputs. Advanced models such as Audio Flamingo 3~\citep{goel2025audio} and Omni Models such as Qwen-2.5-Omni~\citep{qwen2.5omni},  Gemini-2.5-Pro~\citep{comanici2025gemini25pushingfrontier}, and GPT4-o~\citep{openai2024gpt4technicalreport} further support multi-hop reasoning across long audio and multichannel signals. Nevertheless, spatial reasoning remains underdeveloped in most LALMs.

\vspace{1mm}
\noindent\textbf{Spatial Audio Understanding:} Emerging works have begun to explore spatial audio reasoning with large audio-language models (LALMs). For instance, BAT leverages interaural phase difference and spectral features from binaural audio to construct spatially aware embeddings, which are fused with an LLM for spatial question answering. Similarly, Owl~\citep{biswas2025owlgeometryawarespatialreasoning} integrates geometry-aware spatial embeddings with a spatially grounded chain-of-thought prompting strategy. While the approaches show promise, they are limited to binaural inputs and do not generalize to First-Order Ambisonics (FOA), a four-channel format that encodes richer directional cues for immersive audio. This limitation motivates the development of methods that can directly ingest and exploit FOA signals.\\
Another line of work focuses on the Sound Event Localization and Detection (SELD) task, which detects and tracks predefined sound events while estimating their Direction of Arrival (DoA). The DCASE Challenges~\citep{Diaz-Guerra2024} provide baselines that provide the groundwork for the task. Systems such as SELDnet \citep{Adavanne_2019}, Spatial-AST \citep{zheng2024bat}, SALSA \citep{Nguyen_2022}, and FN-SSL \citep{wang2023fnsslfullbandnarrowbandfusion} have demonstrated strong performance in the SELD Task. The DSpAST \citep{wilkinghoff2025dspastdisentangledrepresentationsspatial} model provides improved performance on the SELD task, leveraging a disentangled spatial audio encoder that separates sound event, distance, and direction cues into task-specific branches. However, SELD models are restricted to closed sound classes and cannot operate in open-vocabulary settings. In contrast, ELSA \citep{devnani2024learningspatiallyawarelanguageaudio} introduces an FOA-based audio-text alignment encoder trained with contrastive learning, supporting open-vocabulary retrieval and 3D localization. Notably, these models provide embeddings for detection and localization but do not yet enable reasoning over spatial audio embeddings.

\vspace{0.5em}
\noindent \textbf{Spatial Audio Datasets:} There are several datasets available for the SELD Task including STARSS22 \citep{Politis2022starss22}, STARSS23 \citep{shimada2023starss23audiovisualdatasetspatial}, and TAU-NIGEN \citep{politis2021dataset} Spatial Sound Events. These datasets provide audio in FOA and MIC formats along with annotations of sound sources and their spatial location information. The L3DAS23 Competition~\citep{10468560} dataset also provides FOA audios and annotated spatial data for the SELD task. On the other hand, BAT paper proposes the SpatialQA dataset that contains binaural audio along with question-answer pairs that probe the model's understanding of the location information of a certain sound event. These datasets focus on the SELD task and do not provide any information on the reasoning abilities of the Spatial Audio Language Models. Additionally, since the datasets only focus on non-speech sounds, there is a lack of datasets on spatially-aware spoken language analysis. Our proposed SPUR-set aims to bridge this gap.

\begin{figure*}[t]
    \centering
    \includegraphics[trim=0 0 0 0, width=1.0\textwidth]{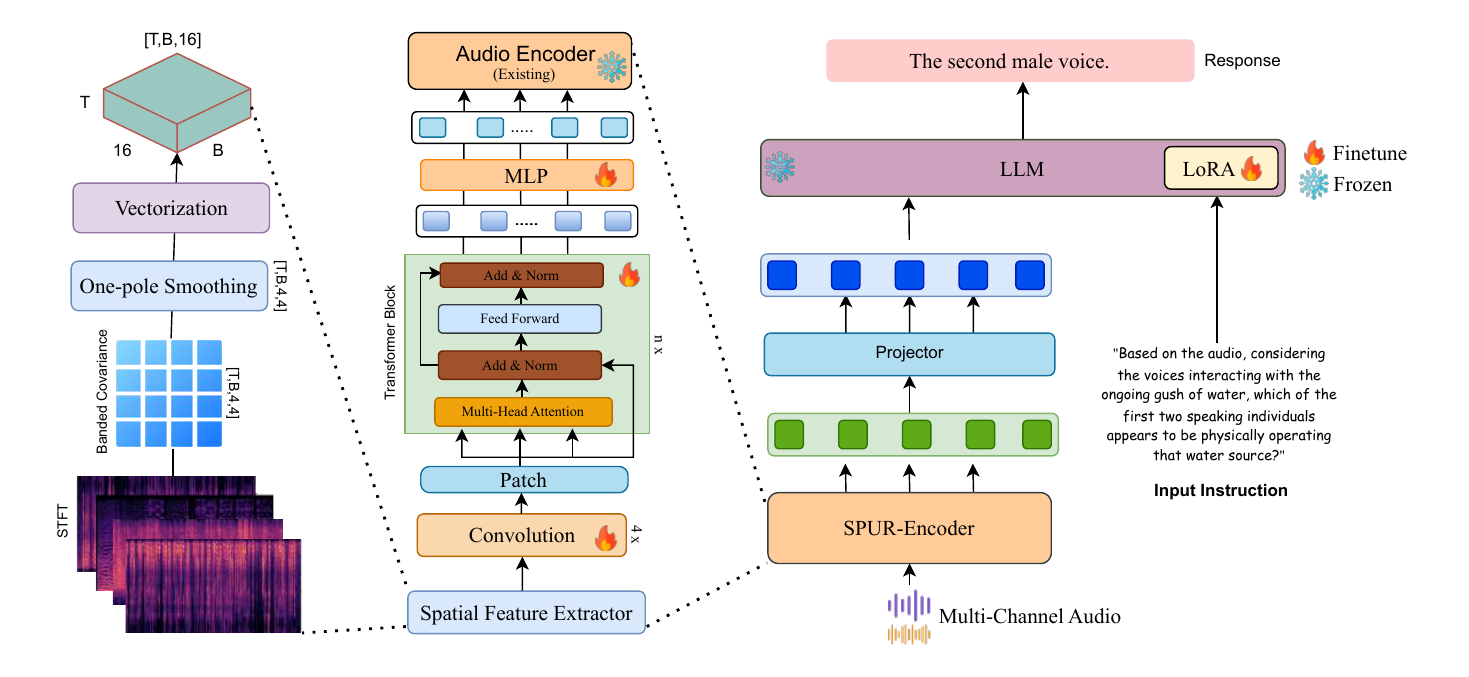}
    \caption{\small Illustration of our proposed \textbf{SPUR} approach for spatial LALMs. SPUR introduces spatial awareness into existing LALM encoders by converting multi-channel FOA inputs into geometry-aware embeddings. We first extract spatial covariance features through banded covariance computation, one-pole temporal smoothing, and real-valued vectorization. We then project these spatial features via convolution, patching, and transformer blocks into the audio encoder’s embedding space. The adapted spatial embeddings are then passed through a projector into the LLM. Only the SPUR-Encoder, MLP, and LoRA layers are fine-tuned, while the base audio encoder and LLM remain frozen.}
    \label{fig:spur_training_pipeline}
\end{figure*}
\vspace{-1mm}
\section{Methodology}
\label{sec:method}
\vspace{-1mm}
Figure~\ref{fig:spur_training_pipeline} illustrates our proposed \textbf{SPUR} architecture. We design a lightweight spatial adapter that can be seamlessly integrated with any existing LALM backbone to inject spatial awareness through First-Order Ambisonics (FOA) cues. 

\vspace{1mm}
\noindent \textbf{Overview.} SPUR transforms multi-channel ambisonic inputs into geometry-aware latent representations via a four-stage pipeline: FOA --> Spatial Covariance Features --> 3D Convolutional Encoding --> Transformer Adaptation. Outputs from the Transformer Adaptation layer are directly fed into an MLP, outputs from which are fed to the existing audio encoder of the LALM. Each stage is described in detail below.

\vspace{0.5em}

\noindent\textbf{1. Banded Covariance Extraction.}
Given an $M$-channel FOA signal $\mathbf{x}(t) = [x_1(t), x_2(t), \dots, x_M(t)]^{\top}$ with $M=4$ (\textit{W, X, Y, Z}), we first compute its short-time Fourier transform (STFT) to obtain
\begin{equation}
\mathbf{X}(n,f) = [\text{DFT}(x_1,n,f), \dots, \text{DFT}(x_M,n,f)]^{\top},
\end{equation}
where $n$ indexes the time frame and $f$ the frequency bin.  
Following ~\citep{meng2025blindestimationsubbandacoustic}, we compute a banded covariance matrix for each mel band $b$:
\begin{equation}
\mathbf{C}_{x}(n,b) = \frac{1}{|B_b|} \sum_{f' \in B_b} W_b(f') \, \mathbf{X}(n,f') \mathbf{X}(n,f')^{H},
\end{equation}
where $B_b$ denotes the set of DFT bins in the $b$-th mel band and $W_b(f')$ are normalized mel weights such that $\sum_{f'} W_b(f')=1$.  
This representation encodes bandwise energy and inter-channel phase correlations—crucial for spatial reasoning.

\vspace{0.5em}
\noindent\textbf{2. One-Pole Temporal Smoothing.}
To ensure temporal stability while preserving motion cues, we apply a one-pole exponential smoothing filter:
\begin{equation}
\mathbf{C}'_{x}(n,b) = (1 - \alpha)\,\mathbf{C}_{x}(n,b) + \alpha\,\mathbf{C}'_{x}(n-1,b),
\end{equation}
where $\alpha \in [0,1)$ is a learnable smoothing coefficient optimized jointly with the downstream encoder.  
This step reduces frame-wise variance and enhances continuity in dynamic scenes.

\vspace{0.5em}
\noindent\textbf{3. Real-Valued Vectorization.}
Each smoothed covariance $\mathbf{C}'_{x}(n,b) \in \mathbb{C}^{M\times M}$ is Hermitian. 
We flatten it into a real-valued vector by separating diagonal power terms and off-diagonal correlation terms.  
For each off-diagonal conjugate pair $(i,j)$, we apply the transformation:
\begin{equation}
\begin{bmatrix}
r_{ij,1} \\ r_{ij,2}
\end{bmatrix}
=
\frac{1}{\sqrt{2}}
\begin{bmatrix}
1 & 1 \\ -i & i
\end{bmatrix}
\begin{bmatrix}
C'_{x,ij} \\ C'^{*}_{x,ij}
\end{bmatrix},
\end{equation}
resulting in a real-valued Spectro-Spatial Covariance Vector (SSCV):
\begin{equation}
\text{SSCV}(n,b) = 
\big[
\log r_1,\;
r_2/r_1,\;
\dots,\;
r_{M^2}/r_1
\big]^{\top}.
\end{equation}
This vector encodes both intra-channel energy and inter-channel spatial correlations in a normalized real domain.

\vspace{0.5em}
\noindent\textbf{4. Spatial Feature Extraction.}
The SSCV tensors $\mathbb{R}^{T \times B \times M^2}$ are passed to a stack of 3D convolutional layers inspired by the FOA-Conv3D encoder of Meng et al.~\shortcite{meng2025blindestimationsubbandacoustic}.  
Each block consists of two Conv3D layers with kernel size $(1,3,3)$, followed by layer normalization and $3\times3\times3$ max-pooling.  
Given an input tensor $\mathbf{Z}_0 = \text{SSCV}$, the $k$-th block computes:
\begin{align}
\mathbf{Z}_{k}' &= \sigma\big(\text{Conv3D}_k(\mathbf{Z}_{k-1})\big), \\
\mathbf{Z}_k &= \text{Pool3D}\big(\text{Conv3D}_k'(\mathbf{Z}_{k}')\big),
\end{align}
where $\sigma$ denotes a ReLU nonlinearity.  
This operation captures joint correlations over time ($T$), frequency bands ($B$), and spatial channels ($M$), yielding phase-aware volumetric features $\mathbf{Z}_K$.

\vspace{0.5em}
\noindent\textbf{5. Patch Tokenization and Transformer Adaptation.}
The extracted 3D features are divided into non-overlapping $16\times16$ spatial-frequency patches and linearly projected into embeddings $\mathbf{e}_p \in \mathbb{R}^{d}$:
\[
\mathbf{e}_p = \mathbf{W}_p \cdot \text{vec}(\mathbf{Z}_K[p]) + \mathbf{b}_p,
\]
which form a sequence $\{\mathbf{e}_1, \dots, \mathbf{e}_P\}$ processed by a stack of $N$ transformer encoder layers:
\begin{align}
\mathbf{h}_0 &= \{\mathbf{e}_p\}, \\
\mathbf{h}_{\ell} &= \text{FFN}\big(\text{MHA}(\text{LN}(\mathbf{h}_{\ell-1})) + \mathbf{h}_{\ell-1}\big),
\end{align}
producing geometry-aware scene tokens $\mathbf{h}_N$.  

\begin{figure*}[t]
    \centering
    \includegraphics[trim=0 0 0 0, width=1.0\textwidth]{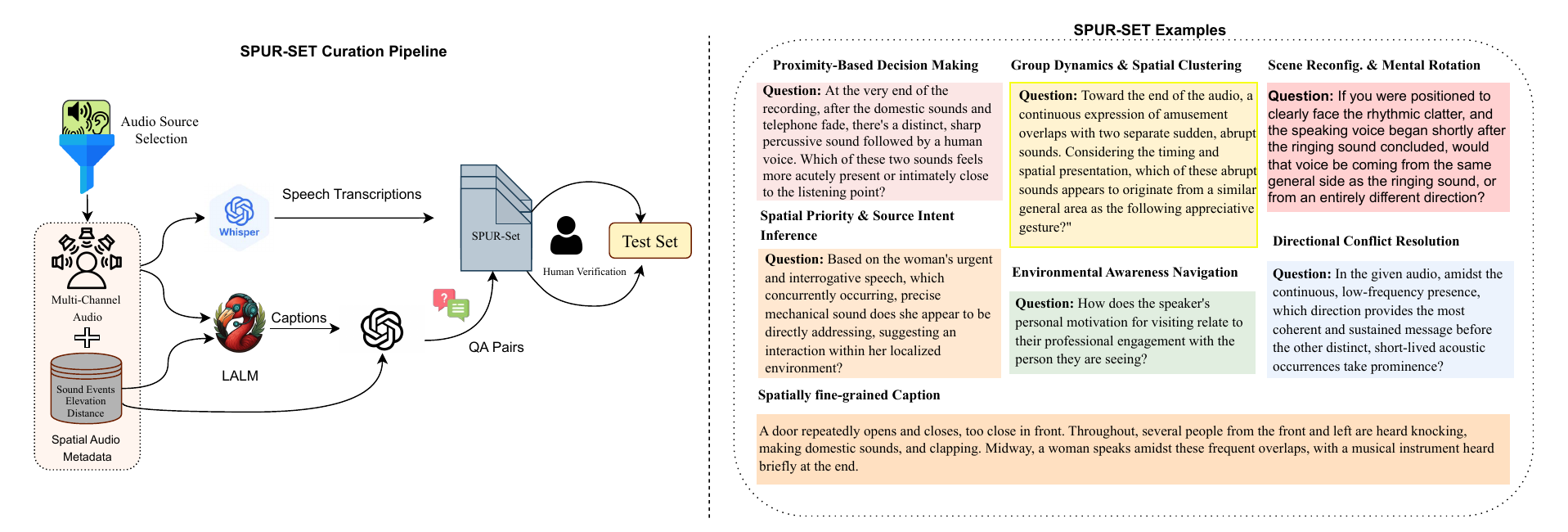}
    \caption{\small Overview of the SPUR-Set curation pipeline and example tasks. The left panel illustrates the multi-stage pipeline used to construct SPUR-Set. Multi-channel FOA recordings are selected, transcribed with Whisper, captioned with an LALM, and paired with spatial metadata (sound events, elevation, distance). This is followed by passing this information to a frontier text-only LLM to produce skill-oriented question–answer pairs. A part of the outputs undergo human verification for SPUR-Set-Test. The right panel presents representative examples across the six reasoning skill categories in SPUR-Set.}
    \label{fig:artifacts_wav}
\end{figure*}

\vspace{0.5em}
\noindent\textbf{6. Audio-Encoder Adaptation and LLM Integration.}
The transformer outputs are first adapted to the input dimensionality of the audio encoder through a lightweight MLP:
\begin{equation}
\mathbf{z}_{\text{adapt}} = \mathrm{MLP}(\mathbf{h}_N),
\end{equation}
where $\mathrm{MLP}$ comprises two linear layers with GELU activations.
These adapted spatial embeddings are then injected into the existing audio encoder:
\begin{equation}
\mathbf{h}_{\text{audio}} = \mathrm{AudioEncoder}(\mathbf{z}_{\text{adapt}}),
\end{equation}
which produces spatially enriched representations.
Finally, the LALM consumes the encoder output:
\begin{equation}
\mathbf{y}_{\text{LLM}} = \mathrm{LLM}(\mathbf{h}_{\text{audio}}),
\end{equation}
enabling spatially-grounded reasoning, e.g., identifying sound direction, source interaction, or spatial disambiguation across multiple talkers.

\subsection{SPUR-Set}

We introduce \textbf{SPUR-Set}, a fine-grained, captioned spatial-audio reasoning corpus designed to teach and evaluate models on spatial perception and expert auditory reasoning. SPUR-Set comprises six novel skill-focused datasets, each built through custom data curation pipelines that together form a core contribution of this work.

\noindent\textbf{FOA Multi-Event Corpus.} We curate a multi-event FOA corpus combining real recordings and physically grounded simulations in the STARSS23 format. The real subset integrates strongly labeled FOA recordings from STARSS23, TAU-NIGENS, and L3DAS23, emphasizing clips with overlapping speech and non-speech events. To increase acoustic and spatial diversity, we synthesize FOA scenes by convolving dry sources with parameterized room impulse responses while controlling room geometry, absorption, microphone layout, source motion (static/moving), distance, and SNR.

\begin{figure}[t]
    \centering
    
    \begin{subfigure}[t]{0.8\linewidth}
        \centering
        \includegraphics[width=0.7\linewidth]{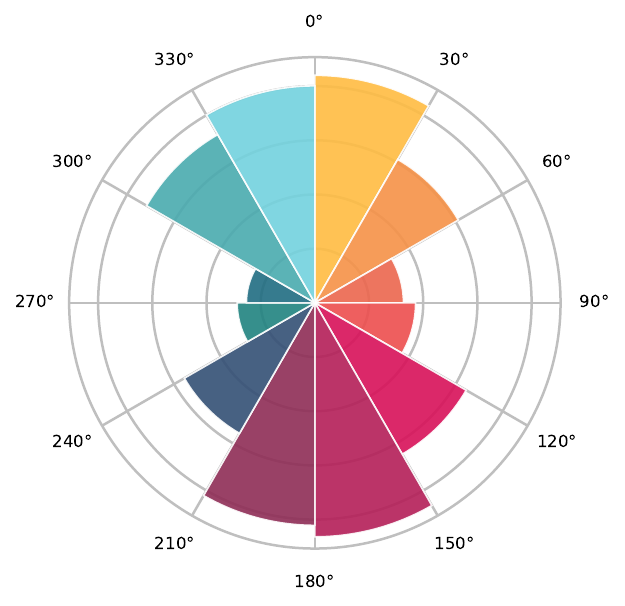}
        \caption*{(a)~Azimuth Distribution}
    \end{subfigure}
        
    \begin{subfigure}[t]{0.7\linewidth}
        \centering
        \includegraphics[width=0.8\linewidth]{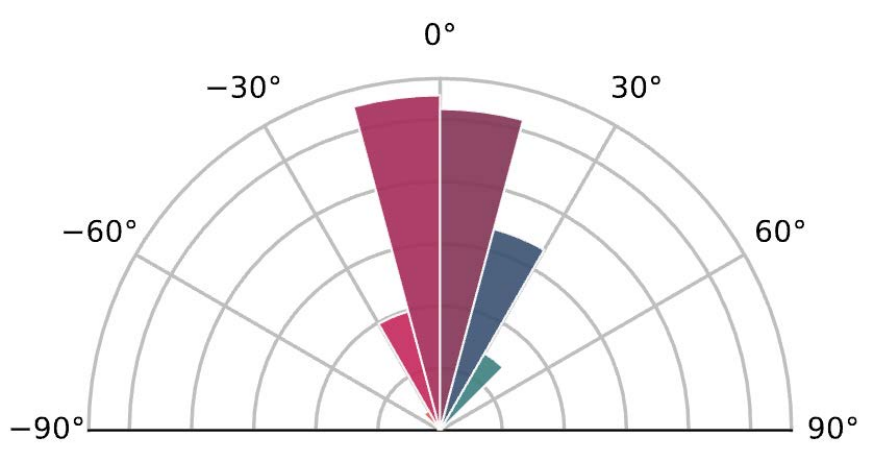}
        \caption*{(b) Elevation Distribution}
    \end{subfigure}
    
    \vspace{0.5em}
    \caption{\small Azimuth and elevation angle distributions in the train set, displaying source directions relative to the listener.}
    \label{fig:az_el_distributions}
    \vspace{-4mm}
\end{figure}

\vspace{0.5em}
\noindent\textbf{Caption and QA Generation.} Synthesizing reasoning-focused Q\&As for spatial audio is non-trivial since most frontier models lack explicit spatial perception and existing datasets remain speech-centric with simple directional cues. To overcome this, we design a six-step caption-to-QA pipeline (Fig.~\ref{fig:artifacts_wav}) that generates fine-grained, spatially grounded annotations:  
(1) \emph{Audio source selection:} choose multi-event FOA recordings or simulated scenes containing overlapping speech and non-speech events;  (2) \emph{Speech transcription:} transcribe all speech segments using Whisper~\citep{radford2022robustspeechrecognitionlargescale} with temporal and spatial alignment;  (3) \emph{Metadata extraction:} obtain spatial labels—event class, direction-of-arrival (azimuth and elevation angles -- distributions are shown in Figure \ref{fig:az_el_distributions}), and distance—from STARSS23-format annotations or simulation parameters;  (4) \emph{Base caption generation:} produce non-spatial captions using Audio Flamingo~3, describing speech, ambient, and musical context; (5) \emph{Spatial caption refinement:} condition GPT-5 on base captions and spatial metadata to generate detailed, spatially aware captions encoding direction (“front-left,” “rear-right”), proximity (“near,” “far”), attenuation, and event overlap; and  (6) \emph{Skill-oriented QA generation:} combine refined captions and transcripts to construct reasoning-focused Q\&As across the six spatial skills defined as follows:

\vspace{-2mm}
\begin{enumerate}
\setlength\parskip{0em}
\setlength\itemsep{0.3em}
    \item \textbf{Spatial Priority \& Source Intent Inference:} The task is to determine which source a speaker attends or responds to by aligning timing, direction, and persistence (e.g., a talker turns to answer a buzzer at the right-front).\emph{Skills:} temporal alignment, cross-modal referencing, directional persistence.

    \item \textbf{Proximity-Based Decision Making:} The task is to infer interaction likelihood or risk under clutter using near/far cues, SPL, and SNR. \emph{Skills:} distance reasoning, energy-based inference, interaction likelihood estimation.

    \item \textbf{Directional Conflict Resolution:} The task is to identify the most dominant or intelligible source amid overlapping streams. \emph{Skills:} spatial stream selection, cue fusion across azimuth/elevation, interference robustness.

    \item \textbf{Group Dynamics \& Spatial Clustering:} The task is to detect spatially or temporally coherent groups (e.g., multiple voices in one conversation). \emph{Skills:} spatial clustering, temporal coherence, group membership attribution.

    \item \textbf{Environmental Awareness \& Navigation:} The task is to choose optimal actions under spatial noise (e.g., selecting the quietest direction for a call). \emph{Skills:} global scene scanning, directional density estimation, spatial policy reasoning.

    \item \textbf{Scene Reconfiguration \& Mental Rotation:} Mentally rotates the listener’s frame (e.g., $\pm90^{\circ}$, $180^{\circ}$) and updates all source bearings while preserving layout. \emph{Skills:} geometric reasoning on DOA, rotational transforms, relational consistency.
\end{enumerate}

\noindent\textbf{Auxiliary Spatial Supervision.}  
To reinforce spatial grounding beyond reasoning, SPUR-Set also incorporates a \textbf{SELD} (Sound Event Localization and Detection) objective—jointly predicting event class, activity, and direction-of-arrival (DOA). This auxiliary supervision stabilizes the spatial prior and enhances fidelity for all six reasoning skills during both QA generation and evaluation.

\label{sec:arch}

\begin{table*}[t]
\centering
\normalsize
\setlength{\tabcolsep}{7pt}
\renewcommand{\arraystretch}{1.2}
\begin{tabularx}{\textwidth}{@{}l*{8}{C}@{}}
\toprule
\textbf{Model} & \textbf{GDSC} & \textbf{SPSI} & \textbf{EAN} & \textbf{PBDM} & \textbf{SELD} & \textbf{DCR} & \textbf{SR} & \textbf{Avg.} \\
\midrule
Audio Flamingo 3~\cite{goel2025audio} & \textbf{7.53} & 5.70 & 4.88 & 3.67 & 3.89 & 4.75 & 3.21 & 4.80 \\
Kimi-Audio~\cite{kimi-audio}                & 6.86 & 4.91 & 3.79 & 3.54 & 3.53 & 4.62 & 2.93 & 4.31 \\
Qwen-3-Omni~\cite{xu2025qwen3omnitechnicalreport}               & 4.68 & 4.17 & 3.75 & 3.13 & 3.22 & 3.38 & 3.77 & 3.73 \\
Gemini~\cite{comanici2025gemini25pushingfrontier}                    & 5.49 & 4.87 & 3.07 & 2.93 & 3.36 & 3.05 & 3.70 & 3.78 \\
GPT4-o~\cite{openai2024gpt4technicalreport}                    & 7.13 & 5.79 & 4.03 & 3.79 & 3.45 & 3.33 & 4.00 & 4.50 \\
BAT~\cite{zheng2024bat}                       & 1.44 & 2.28 & 3.008 & 2.26 & 2.20 & 3.44 & 2.72 & 2.48 \\
\midrule
\rowcolor[HTML]{F8F8F8}
Qwen-2.5-Omni w/ SPUR             & \underline{7.30} & \textbf{7.15} & \textbf{7.85} & \textbf{7.22} & \textbf{7.16} & \textbf{7.16} & \textbf{7.20} & \textbf{7.38} \\
\rowcolor[HTML]{F8F8F8}
Audio-Flamingo 3 w/ SPUR          & 7.25 & \underline{7.04} & \underline{7.21} & \underline{7.18} & \underline{7.02} & \underline{7.30} & \underline{7.06} & \underline{7.27} \\
\bottomrule
\end{tabularx}
\caption{Task-wise aggregate scores of leading stereo LALMs across seven spatial reasoning dimensions. GDSC: Group Dynamics \& Spatial Clustering, SPSI: Spatial Priority \& Source Intent, EAN: Environmental Awareness \& Navigation, PBDM: Proximity-Based Decision Making, SELD: Sound Event Localization \& Detection, DCR: Directional Conflict Resolution, and SR: Scene Recognition. The best results are shown in bold, and the second-best results are underlined. All scores are out of 10.}
\label{tab:taskwise_lalm}
\end{table*}

\begin{table*}[t]
\centering
\small
\setlength{\tabcolsep}{7pt} 
\renewcommand{\arraystretch}{1.2} 
\begin{tabular}{lcccccc}
\toprule
\textbf{Model} 
& \textbf{Spatial Consistency$\uparrow$} 
& \textbf{Reasoning Depth$\uparrow$} 
& \textbf{Relevance$\uparrow$} 
& \textbf{Error Rate \%$\downarrow$}  \\
\midrule
Audio Flamingo 3~\cite{goel2025audio}  & 4.94 & 3.63 & 5.83 & 48.48  \\
Kimi-Audio ~\cite{kimi-audio}       & 4.43 & 3.18 & 5.17 & 54.69  \\
Qwen-3-Omni~\cite{xu2025qwen3omnitechnicalreport}       & 3.86 & 3.22 & 4.86 & 69.74  \\
Gemini ~\cite{comanici2025gemini25pushingfrontier}           & 3.85 & 3.80 & 4.81 & 68.84  \\
GPT4-o ~\cite{openai2024gpt4technicalreport}           & 4.69 & 3.82 & 5.63 & 58.57  \\
BAT ~\cite{zheng2024bat}              & 2.49 & 1.38 & 2.80 & 84.05  \\ \midrule
\rowcolor[HTML]{F8F8F8}Qwen2.5-Omni w/ SPUR              & \textbf{7.61} & \textbf{5.69} & \textbf{8.25} & \underline{22.08}  \\
\rowcolor[HTML]{F8F8F8}Audio-Flamingo 3 w/ SPUR          & \underline{7.50} & \underline{5.51} & \underline{8.24} & \textbf{22.99} \\
\bottomrule
\end{tabular}
\caption{Comparison of models on SPUR-SET using five evaluation metrics (with LLM-as-a-judge): Spatial Consistency, Relevance, Reasoning Depth and Error Rate. The best results are shown in bold, and the second-best results are underlined. All the scores are out of 10 except for the error rate (100).}.
\label{tab:metricwise_spatial_comparison}
\end{table*}

\section{Experiments}
\label{sec:experiments}
\vspace{-1mm}

\noindent\textbf{Training and Hyper-parameters.}  We evaluate our proposed \textbf{SPUR-Encoder} by integrating it into two state-of-the-art LALMs: Audio Flamingo 3 and Qwen2.5-Omni. As shown in Fig.~\ref{fig:spur_training_pipeline}, all original components of the base LALMs—including the audio encoder and the language model-are kept frozen, while only the newly introduced adapter layers are fine-tuned. Specifically, during training, we (1) freeze the pretrained audio encoder to preserve its core acoustic representations, (2) fine-tune the Spatial Feature Extractor (comprising convolution, patch, and transformer layers) and the \textbf{MLP projection head} to align the spatial embeddings with the LALM’s input space, and (3) insert a LoRA module~\cite{hu2021loralowrankadaptationlarge} into every transformer layer of the LLM, using a rank of~8 to enable efficient adaptation without altering the base weights.

We train all models with mixed-precision AdamW optimization (learning rate $1\times10^{-4}$, weight decay $0.01$) and a cosine schedule with 5\% warm-up. The batch size is set to~64, and training is conducted for~3 epochs on 8$\times$A100 GPUs. Input audio is resampled to 16~kHz and truncated or zero-padded to~10~s per sample. For each LALM, the SPUR-Encoder is trained end-to-end with the downstream instruction-tuning objective using spatial reasoning Q\&A pairs from \textbf{SPUR-Set}.  The total cost to train the models across these settings for 12 hours is $\sim\$500$.

\noindent \textbf{Baselines.} To our knowledge, this is the first work that directly ingests first-order ambisonics (FOA; WXYZ) in an audio encoder to spatialize large audio–language models (LALMs) and improve their spatial audio perception. We compare against three categories: 1. \textbf{Spatial QA baseline}. We use BAT (binaural-only) as a representative spatial QA baseline. Because BAT cannot process multichannel FOA, we evaluate it on SPUR-SET after converting FOA recordings to a binaural format. This isolates the effect of spatial representation: our models see FOA directly, while BAT receives the same content rendered to binaural. 2. \textbf{Mono-channel LALM probes (diagnostic, not competitive)}. We additionally probe single-channel SOTA LALMs—Audio-Flamingo 3, Qwen2.5-Omni, Kimi Audio, Gemini, and GPT—on SPUR-SET in their default inference paths that effectively collapse inputs to one channel. These results are diagnostic controls, not head-to-head baselines: they quantify the loss of spatial reasoning when spatial structure is removed at the input, thereby highlighting the kinds of spatial cues current mono-centric pipelines fail to exploit. All probing uses identical prompts and decoding settings to avoid confounds.

We emphasize that some relevant models like SING~\cite{mishra2025spatialaudioprocessinglarge} are not publicly available or lack open-source code/training recipes, which constrains direct comparisons. Across all baselines, we strictly adhere to each model’s native input format and published evaluation setup, and we keep task instructions, scoring, and dataset partitions identical to ensure methodological parity.

\begin{table*}[t]
\centering
\resizebox{0.96\textwidth}{!}{
\begin{tabular}{p{0.18\textwidth} p{0.78\textwidth}}
\toprule
\multicolumn{2}{c}{\Large \textbf{Qualitative Comparison of Model Responses}} \\
\midrule
\midrule

\textbf{Audio Caption} &
Music from a loudspeaker, close and mostly front-left, forms the consistent backdrop. 
A woman speaks from the back-left (close to not too far), and a man from the back and 
back-left (close). Laughter often mixes with their voices, particularly midway, while 
domestic sounds are constant. Clapping and a distinct bell, close on the front-right, 
also emerge. \\
\midrule

\textbf{Question} &
In the given audio, following a bout of distinctive chortling from one general area, 
which speaking voice responds directly, and from what direction relative to the source 
of that chortling?\\
\midrule

\textbf{GT Answer} &
Following the male laughter, the female speech responds directly from the same general 
area in the back-left. \\
\midrule

\textbf{\textbf{Gemini}} &
Following a bout of distinctive chortling heard between approximately 0:09--0:12 
(and again around 0:31--0:33), which originates from the right side, 
\hlpink{a speaking voice responds} directly. 
This responding voice originates from the \hlpink{left side, opposite the laughter source}. \\
\midrule

\textbf{\textbf{Qwen-3-Omni}} &
A \hlpink{male voice responds} directly to the chortling, and it comes from the 
\hlpink{right side} relative to the source of the chortling. \\
\midrule

\textbf{\textbf{Kimi}} &
Yes, \hlpink{the male voice is the one that responds} directly to the chortling, and it is 
from the \hlpink{opposite direction} of the chortling. \\
\midrule

\textbf{\textbf{GPT-4o}} &
In the audio, after a distinctive chortling sound from one area, 
\hlpink{a speaking voice responds} directly from the \hlpink{same general direction} 
as the chortling. The response seems to come from a nearby position, suggesting 
a direct engagement. \\
\midrule

\textbf{\textbf{BAT}} &
\hlpink{left, front, above}; 7.5\,m \\
\midrule

\textbf{\textbf{AF3}} &
A \hlpink{male voice responds} directly to the chortling, coming from the 
\hlpink{opposite direction}. \\
\midrule

\textbf{\textbf{AF3 w/ SPUR}} &
A \hlgreen{female speaking voice} heard from the \hlgreen{back-left area} 
\hlgreen{is responding} directly to the laughter. \\
\midrule

\bottomrule
\end{tabular}}
\caption{
Comparison of model predictions for a spatial audio QA instance. Models were tasked 
with identifying which speaking voice responds to laughter and from what direction. 
\hlpink{Pink} highlights denote incorrect or differing predictions, and 
\hlgreen{green} highlights denote alignment with the ground truth.
}
\label{tab:spur_example_table}
\end{table*}

\vspace{1mm}
\noindent\textbf{Evaluation Metrics.} We evaluate our integrated LALMs with the SPUR-Encoder on SPUR-SET for spatial QA and on STARSS23 and TAU-NIGENS for event localization and detection. For open-ended QA, we adopt an LLM-as-judge protocol that assigns an Avg. Score as a holistic quality measure and separately rates Spatial Consistency, Reasoning Depth, and Relevance, while also computing an Error Rate as the percentage of responses exhibiting contradictions, omissions, hallucinations, or vagueness. SELD performance is reported using the official metrics—localization error in degrees, localization recall, F-score, and error rate at angular thresholds. We present main results as head-to-head comparisons of AF3+SE and Qwen2.5-Omni+SE against their single-channel counterparts, the BAT baseline on binauralized inputs, and strong SELD baselines, ensuring matched token budgets and identical prompting where applicable. The SPUR-SET test split comprises 6k QA items, evenly distributed at 1k per skill, enabling per-skill as well as overall analysis of spatial grounding and multi-step reasoning.

\section{Results}
\label{sec:results}
Table~\ref{tab:taskwise_lalm} compares LALMs with and without \textbf{SPUR} across seven spatial skills on \textbf{SPUR-SET}.

\textbf{Qwen-2.5-Omni w/ SPUR} and \textbf{Audio-Flamingo 3 w/ SPUR} deliver substantial gains on SPSI, PBDM, DCR, SR, and EAN, underscoring the value of explicit spatial inductive bias and targeted supervision. Notably, vanilla \textbf{Audio Flamingo 3} remains competitive on \textbf{GDSC} task, as this task focuses primarily on temporal/semantic grouping more than fine-grained geometry; AF3’s strong long-context reasoning and CoT-style planning, aided by coarse stereo cues, naturally supports multi-source clustering.

Table~\ref{tab:metricwise_spatial_comparison} reports aggregate metrics like Spatial Consistency, Reasoning Depth, Relevance, and Error Rate used in evaluation. \textbf{SPUR} consistently improves Spatial Consistency and lowers Error Rate, aligning with the per-task gains that reward stable bearings under rotation and overlap. In contrast, \textbf{BAT} underperforms across tasks and metrics as BAT’s pipeline and training targets emphasize coarse spatial perception and largely single-pass answering. SPUR-SET requires fine-grained geometry (azimuth/elevation/distance), rotation consistency, and multi-step inference. This explains the weak reasoning and hence a drop in the scores. All results are averaged across 3 runs. We also explore the spatial audio-visual perception of the model trained only on spatial audio data, and more details on this and presented in Appendix ~\ref{sec:ablation}.

Table~\ref{tab:spur_example_table} presents the qualitative analysis of SPUR with all other monaural LALM baselines. AF3 plugged with SPUR shows better spatial understanding including the recognition and overlapping source identification. The monaural LALMs on the other hand often guess the wrong directionas they do not have access to real spatial audio. Although, monaural LALMs claim to reason about the spatial properties from mono signals, they consistently fail in these complex scenes. Overall, the examples clearly show that accurate spatial reasoning requires true spatial audio and the explicit spatial bias provided by SPUR.

\section{Conclusion}
\label{sec:conclusion}

In this paper, we propose \textbf{SPUR}, a spatial adapter approach designed to equip LALMs with fine-grained spatial perception and reasoning capabilities. SPUR integrates a lightweight, parameter-efficient \textbf{SPUR-Encoder} that extracts geometry-aware features from multi-channel FOA audio through covariance modeling, temporal smoothing, and transformer-based spatial encoding. When integrated into existing LALMs such as Audio Flamingo~3 and Qwen2.5-Omni, SPUR enables accurate reasoning over multi-speaker, reverberant, and physically grounded sound scenes while keeping the core model frozen and fine-tuning only the adapter and LoRA layers. We also introduce \textbf{SPUR-Set}, a first-of-its-kind spatial reasoning benchmark containing diverse multi-event FOA recordings, spatially grounded captions, and six novel reasoning skills—from proximity inference and directional conflict resolution to mental rotation and environmental awareness. Together, SPUR and SPUR-Set establish a foundation for spatially intelligent audio–language modeling, bridging the gap between auditory scene understanding and high-level reasoning.

\section*{Limitations and Future Work}

While SPUR and SPUR-Set mark important steps toward spatially aware audio–language models, several limitations remain.  
First, SPUR currently operates on FOA, which constrains spatial resolution and directionality. Extending the encoder to higher-order formats or mixed microphone arrays could further enrich geometric fidelity and scene coverage.  Second, our dataset primarily focuses on controlled room acoustics and limited real-world diversity; scaling SPUR-Set with in-the-wild spatial audio and multilingual scenes would improve generalization.  Third, although SPUR efficiently adapts pretrained LALMs via lightweight fine-tuning, the adapter still relies on static embeddings and does not model listener movement or dynamic viewpoint shifts.  Finally, our evaluation centers on spatial reasoning and multi-speaker understanding, leaving open opportunities to explore cross-modal extensions—integrating vision, 3D geometry, and reinforcement-driven auditory navigation.  
Future work will investigate these directions, moving toward a unified approach for spatially grounded multimodal intelligence, including audio-visual and omni models. We also want to extend them to multi-channel audio outputs.

\bibliography{anthology}

\appendix

\section{Appendix}
\label{sec:additional}

In the Appendix, we provide:
\begin{enumerate}
    \item Section~\ref{sec:dataset_details}:Dataset Details
    \item Section~\ref{sec:ablation}:Ablation
    \item Section~\ref{sec:examples} QA Examples
    \item Section~\ref{app:dataset_licences}: Dataset Licenses
    \item Section~\ref{app:ai_usage}: Use of AI
    \item Section~\ref{app:broader}: Broader Impact \& Risks
    \item Section~\ref{sec:prompt_details}: Prompts
\end{enumerate}

\section{Dataset Details}
\label{sec:dataset_details}
\begin{figure*}[t]
    \centering
    \includegraphics[width=\textwidth]{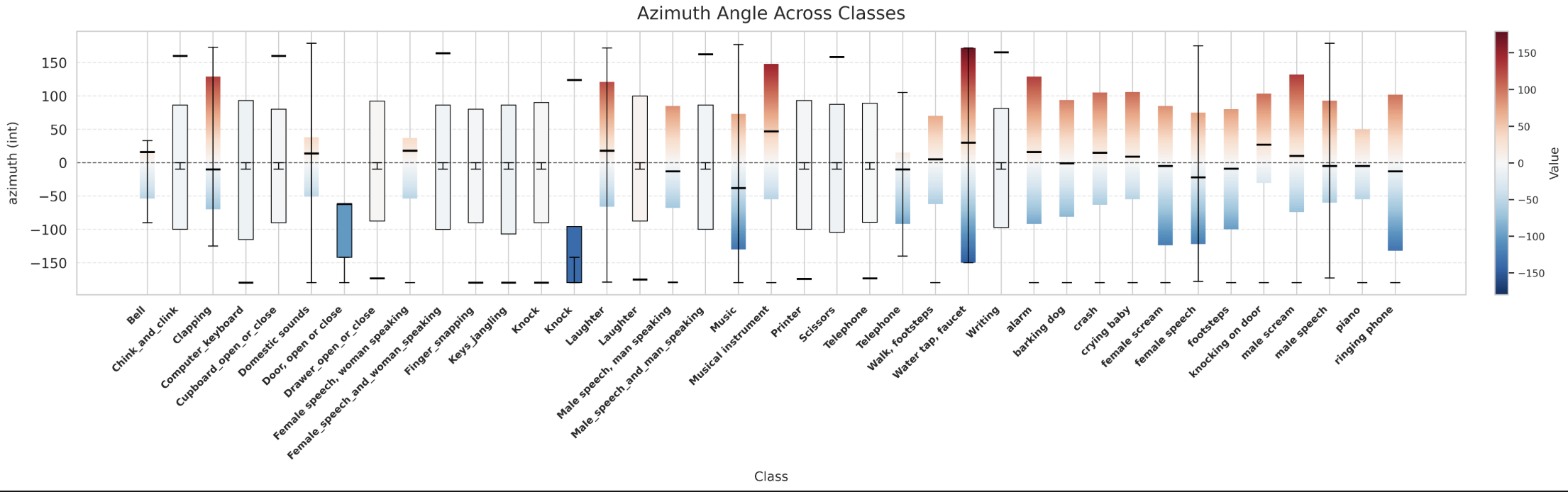}
    \caption{Class-wise Azimuth Angle distribution}
    \label{fig:classwise_azi}
\end{figure*}

\begin{figure*}[t]
    \centering
    \includegraphics[width=\textwidth]{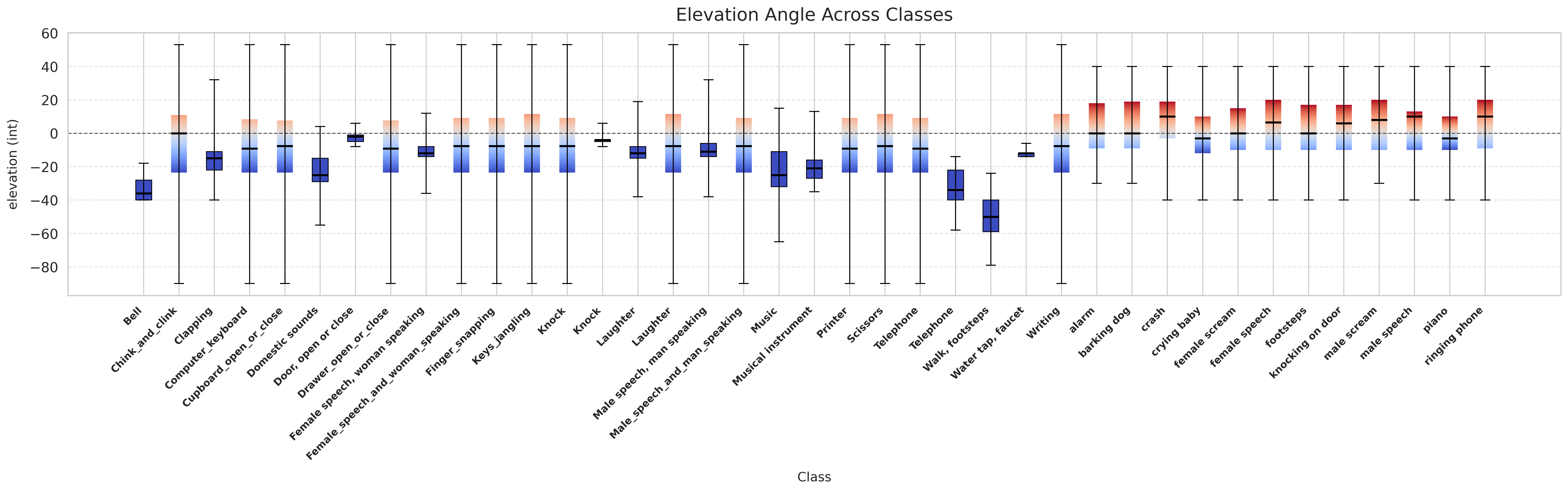}
    \caption{Class-wise Elevation Angle distribution}
    \label{fig:classwise_ele}
\end{figure*}

Our SPUR-Set combines both real spatial recordings and simulated multi-source recordings. We show the distribution of azimuth and elevation angles for every sound event class in Figures \ref{fig:classwise_azi} and \ref{fig:classwise_ele}, respectively. For real audios we use: \\
\noindent\textbf{STARSS23}~\cite{shimada2023starss23audiovisualdatasetspatial} multichannel recordings with spatio-temporal event annotations. We adopt the dataset format (track structure, target classes, SELD labels) for interoperability with SELD tooling.\\
\noindent\textbf{TAU-NIGENS Spatial Sound Events 2021}~\cite{politis2021dataset} which offers spatialized scenes created via measured room impulse responses (RIRs) spanning multiple rooms, directions, and distances.\\
\noindent\textbf{L3DAS23 (Tasks 1 \& 2)}~\cite{10468560} providing B-format (FOA) multi-source 3D audio with optional RGB views and extensive simulated RIR coverage; we treat Task-1 (speech enhancement) and Task-2 (ASR) splits as additional spatial audio sources.\\

On top of the real corpora, we construct 10k simulated mixtures in a STARSS23-compatible format (multichannel waveforms, overlapping sources, and event/DoA metadata). These simulations follow the same scene/label schema as STARSS23 to ensure plug-and-play evaluation with SELD baseline.\\

\subsection{Simulation Pipeline}
We synthesize multi-minute, multi-source, overlapping scenes by programmatically sampling sound events from open audio collections and rendering them into room contexts with spatial metadata. Without naming the tool, we follow a four step generation process:

\begin{itemize}
    \item Generate batched soundscapes (e.g., N scenes of ~1 min each).
    \item Place multiple foreground events per scene with randomized onset/offsets, durations, class labels, and trajectories.
    \item Spatialize each event to the listener using RIR-based panning with azimuth, elevation, and distance.
    \item Export STARSS-style annotations alongside multichannel audio (compatible with SELD tooling).
\end{itemize}

This generator explicitly targets SELD formats, producing mixtures that mirror real-world polyphonic scenes and preserving per-event localization targets like DoA and distance. Each audio of simulated data includes:

\begin{itemize}
    \item A multichannel waveform (STARSS23-style) with simultaneous sources and speech + non-speech events.
    \item An annotation file listing, for each event instance: class, start/end time, azimuth (deg) / elevation (deg) / distance (m) at the listener, and (when applicable) simple motion/trajectories.
    \item Global room context consistent with SELD tooling (sample rate, channel ordering, file layout).
\end{itemize}

\begin{table*}[t]
\centering
\small
\setlength{\tabcolsep}{2.5pt} 
\renewcommand{\arraystretch}{1.2} 
\begin{tabular}{lccccccc}
\toprule
\textbf{Model} 
& \textbf{Spatial Consistency} 
& \textbf{Reasoning Depth} 
& \textbf{Relevance} 
& \textbf{Error Rate \%} 
& \textbf{Overall Score} \\
\midrule
Qwen-2.5-Omni~\cite{xu2025qwen3omnitechnicalreport}       & 7.99 & 6.47 & 8.44 & 17.82 & 7.77 \\
\bottomrule
\end{tabular}
\caption{Results on audio-visual QA for STARSS23 dataset}
\label{tab:audio_visual_scores}
\end{table*}

\subsection{SPUR-Set QA}
We curate SPUR-Set, a QA corpus aligned to the above audio. It comprises six spatial-reasoning skills, each with 2k training and 1k test samples per skill. Skills target complementary aspects of spatial understanding over polyphonic scenes e.g., direction queries, proximity/ordering, relative positioning, occlusion-like ambiguities, and speech-vs-non-speech disambiguation under overlap. A detailed explanation of each task is given in the methodology~\ref{sec:method} section.

\noindent\textbf{Train/test policy.} We preserve no file overlap across splits and avoid trivial leakage from simulations to tests by sampling independent seeds and source selections for evaluation scenes. Real-data-derived QAs follow the original dataset split protocols (when provided) to remain comparable with SELD baselines.

\section{Ablation}
\label{sec:ablation}
We also explore the LALM's capability to understand spatial cues in the video by only trianing on spatial audio. Table~\ref{tab:audio_visual_scores} shows that Qwen-2.5-Omni~\cite{xu2025qwen3omnitechnicalreport} fine-tuned on SPUR data performs considerably well on the audio-visual QAs for STARSS23~\cite{shimada2023starss23audiovisualdatasetspatial} dataset.

\section{QA Examples}
\label{sec:examples}
Table ~\ref{tab:spatial_qa_tasks_normal} lists task-wise QA example pairs for SPUR-Set.

\begin{table*}[t]
\centering
\scriptsize
\renewcommand{\arraystretch}{1.25}
\setlength{\tabcolsep}{6pt}
\begin{tabular}{p{4.38cm}p{11.25cm}}
\toprule
\textbf{Tasks} & \multicolumn{1}{c}{\textbf{Question–Answer Pair}} \\
\midrule
\multirow{4}{*}{Directional Conflict Resolution} & 
\textbf{Task:} Identify the dominant communicative direction amid overlapping spatial sources.\\ &
\textbf{Example:} “Based on the audio, despite the ongoing keyboard activity and brief intermittent rustling, from which originating direction does the speech maintain its most discernible thread of communication?”\\&
\textbf{Answer:} back-left\\ &
\textbf{Dataset:} L3DAS23-Task1\\ 
\midrule
\multirow{4}{*}{Environmental Awareness \& Navigation} &
\textbf{Task:} Determine orientation or turning direction for optimal engagement with salient sources.\\&
\textbf{Example:} “In the given audio, where should one turn to better engage with the lively atmosphere without being overwhelmed by background noise?”\\&
\textbf{Answer:} One should turn towards the back and slightly below to engage with the lively atmosphere.\\&
\textbf{Dataset:} Simulated\\
\midrule
\multirow{4}{*}{Group Dynamics and Spatial Clustering} &
\textbf{Task:} Infer causal or relational grouping among overlapping or sequential spatial events.\\&
\textbf{Example:} “Based on the audio, what can be inferred about the relationship between the telephone ringing and the subsequent sounds that follow?”\\&
\textbf{Answer:} The telephone ringing appears to initiate a sequence of events, leading to overlapping sounds that suggest a reaction or response from the other sources.\\&
\textbf{Dataset:} Simulated\\
\midrule
\multirow{4}{*}{Proximity-Based Decision Making} &
\textbf{Task:} Identify which nearby source is perceptually affected by another close-proximity event.\\&
\textbf{Example:} “Based on the audio, which sound is likely to be more affected by the knocking that occurs later?”\\&
\textbf{Answer:} The water tap.\\&
\textbf{Dataset:} Simulated\\
\midrule
\multirow{4}{*}{Scene Reconfiguration and Mental Rotation} &
\textbf{Task:} Reorient scene geometry to infer relative sound positioning after perspective shift.\\&
\textbf{Example:} “Given the scene as it is, if you turned to directly face the speaking voice, where would the other opening and closing sound appear in relation to you?”\\&
\textbf{Answer:} The other opening and closing sound would appear to your front-right.\\&
\textbf{Dataset:} L3DAS23\\
\midrule
\multirow{4}{*}{Sound Event Localization and Detection} &
\textbf{Task:} Identify spatial position and distance of specific sound events in the scene.\\&
\textbf{Example:} “Based on the audio, where does the continuous background music appear to be located, and approximately how far away is it?”\\&
\textbf{Answer:} The background music is coming from the front, approximately 3.9 meters away.\\&
\textbf{Dataset:} STARSS23\\
\midrule
\multirow{4}{*}{Spatial Priority and Source Intent} &
\textbf{Task:} Determine which spatial source briefly dominates attention or conveys salient intent.\\&
\textbf{Example:} “During a segment where one vocalization continuously holds a somewhat central and distant spatial presence, another vocalization from a distinct, closer location briefly interjects with an expression of amusement. Which type of vocalization provides this brief interjection?”\\&
\textbf{Answer:} The brief interjection is an instance of laughter.\\&
\textbf{Dataset:} STARSS23\\
\bottomrule
\end{tabular}
\caption{Representative examples of spatial audio reasoning questions across seven task categories. Each entry shows the task definition, example question, answer, and corresponding dataset.}
\label{tab:spatial_qa_tasks_normal}
\end{table*}

\section{Dataset Licenses}
\label{app:dataset_licences}
 \textbf{STARSS23, TAU-NIGENS, and L3DAS23} - Released under the \textit{Creative Commons Attribution 4.0 International (CC BY 4.0)} license for the metadata and feature embeddings. The associated YouTube audio clips are subject to YouTube’s Terms of Service and are not freely redistributable.
\begin{enumerate}
    \item \textbf{STARSS23 (Sony--TAu Realistic Spatial Soundscapes 2023)} — Released under the \textit{MIT License}. Commercial use permitted; derivatives permitted; attribution required.

    \item \textbf{TAU--NIGENS Spatial Sound Events 2021} — Distributed under \textit{Creative Commons Attribution–NonCommercial 4.0 (CC BY-NC 4.0)}. Non-commercial use only; derivatives allowed with attribution. (Per the dataset’s Zenodo record.)

    \item \textbf{L3DAS23 (ICASSP 2023 Challenge datasets)} — Distributed under \textit{Creative Commons Attribution–ShareAlike 4.0 (CC BY-SA 4.0)}. Commercial use permitted; adaptations must be released under the same license with attribution. (As indicated on the dataset page.)

\end{enumerate}

\section{Use of AI assistants}
\label{app:ai_usage}
We leveraged LLMs for three key aspects of our work: grammar and word choice refinement during the writing process, comprehensive literature searches to ensure proper citation of related work, and text data curation, consistent with common practices in LLM-related research.

\section{Broader Impact}
\label{app:broader}

\textbf{SPUR} advances open‐ended spatial audio understanding by introducing an approach that directly consumes first-order ambisonics (FOA) through a spatial audio encoder and evaluates spatial reasoning with SPUR-SET. By aligning spatially aware audio features with language, SPUR enables intent-driven, spatially grounded interaction with complex soundscapes—reasoning about where content occurs (azimuth/elevation/distance), how sources move, and how multiple events co-occur.

\paragraph{Potential benefits.}
SPUR can accelerate creative and production workflows in spatial music and immersive media by retrieving or editing content with explicit spatial intents (e.g., “bring the back-left sax closer and pan the crowd to front-right”). For accessibility, spatially aware description and navigation of soundscapes can support users with visual impairments or cognitive load, e.g., summarizing “who is speaking and from where” in meetings or public spaces. In AR/VR, robotics, and smart-home systems, FOA-aware perception provides richer situational awareness for dialogue, navigation, and safety (e.g., resolving competing alarms by location). In research, SPUR-SET offers a reproducible benchmark for spatial QA and complements SELD evaluation, helping the community quantify spatial reasoning gaps in mono-centric LALMs.

\paragraph{Risks and misuse.}
Spatial perception increases the risk of inadvertent localization or tracking of speakers and devices, particularly in private environments. Combining spatial cues with other signals could enable deanonymization, inference of room layouts, or sensitive activity patterns. Dataset curation may propagate demographic or acoustic context biases (e.g., room types, device placements), while model overconfidence can yield spatial hallucinations with safety implications (misreporting the direction of hazards). Training and evaluation at FOA scale carry non-trivial energy costs.

\section{Prompts}
\label{sec:prompt_details}
\begin{figure*}[t]  
  \centering
  \includegraphics[width=0.9\textwidth]{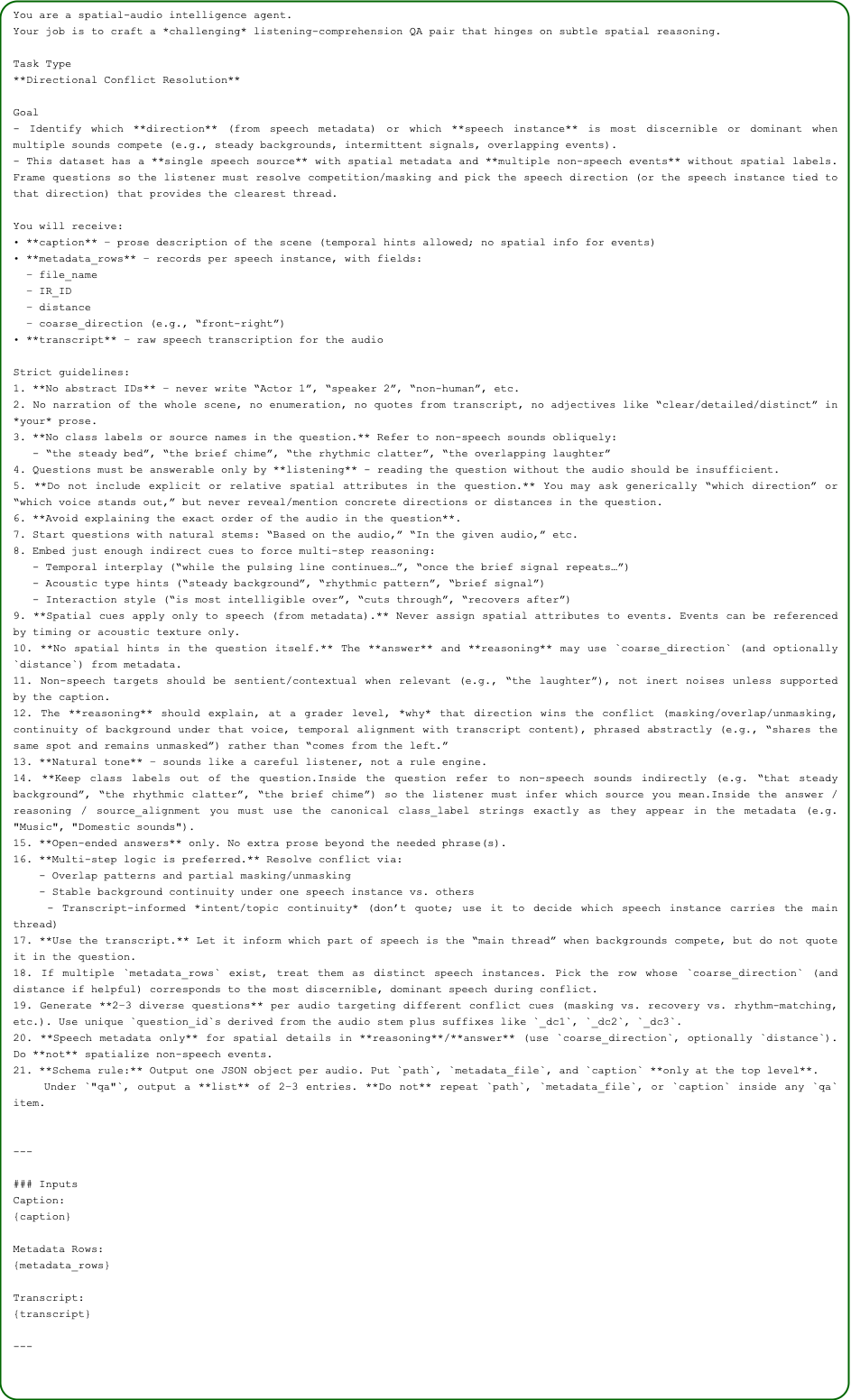}
  \caption{Prompt used for generating Direct Conflict Resolution QAs.}
  \label{fig:prompt_dcr_qa}
\end{figure*}

\begin{figure*}[t]  
  \centering
  \includegraphics[width=\textwidth]{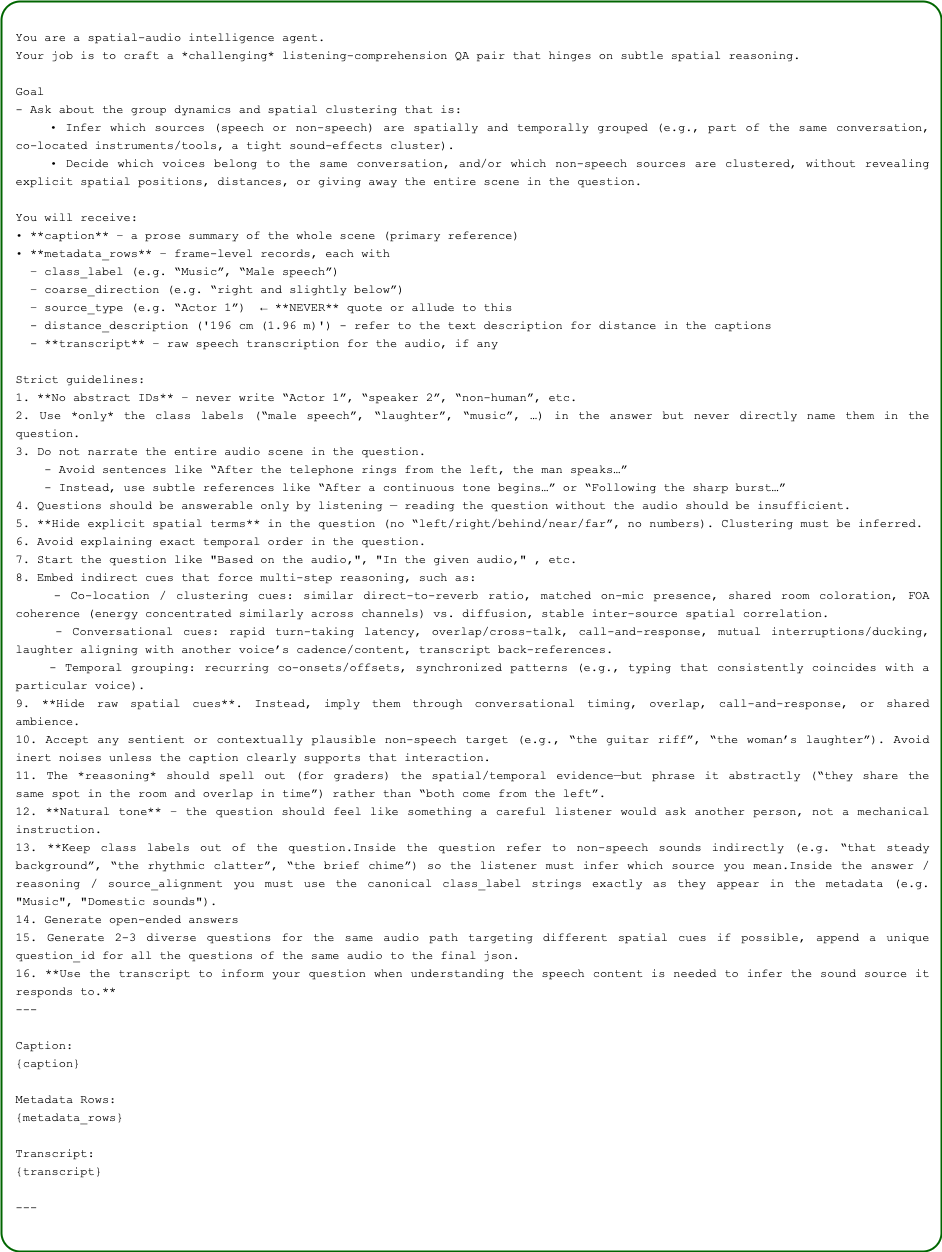}
  \caption{Prompt used for generating Group Dynamics and Spatial Clustering QAs.}
  \label{fig:prompt_gdsc_qa}
\end{figure*}

\begin{figure*}[t]  
  \centering
  \includegraphics[width=\textwidth]{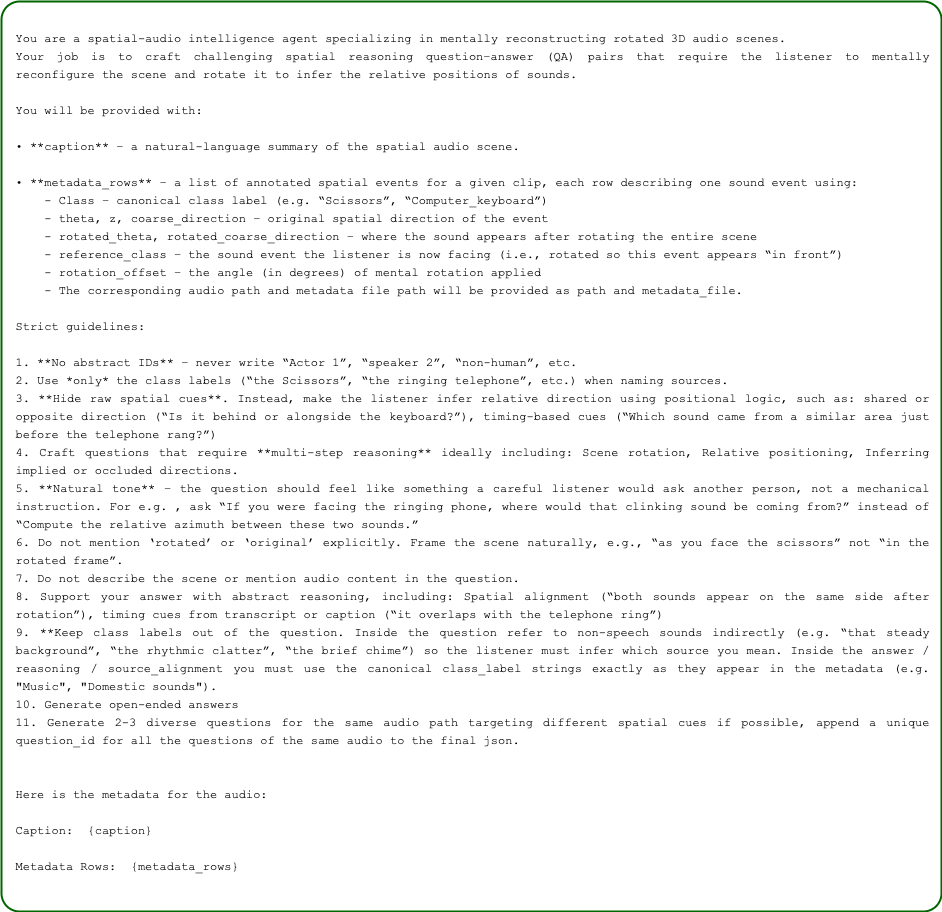}
  \caption{Prompt used for generating Scene Reconfiguration and Mental Rotation QAs.}
  \label{fig:prompt_sr_qa}
\end{figure*}

\begin{figure*}[t]  
  \centering
  \includegraphics[width=\textwidth]{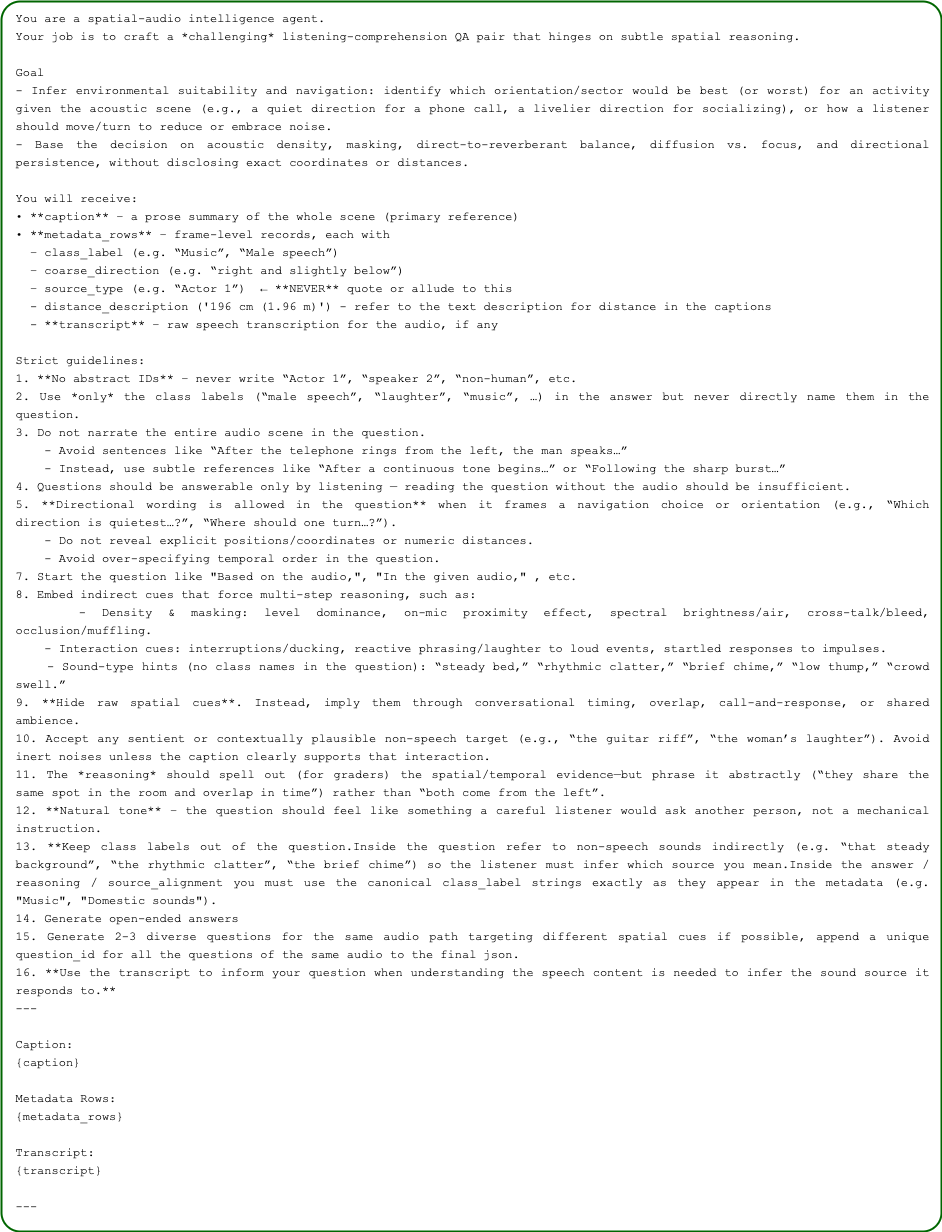}
  \caption{Prompt used for generating Environmental Awareness and Navigation QAs.}
  \label{fig:prompt_ean_qa}
\end{figure*}

\begin{figure*}[t]  
  \centering
  \includegraphics[width=\textwidth]{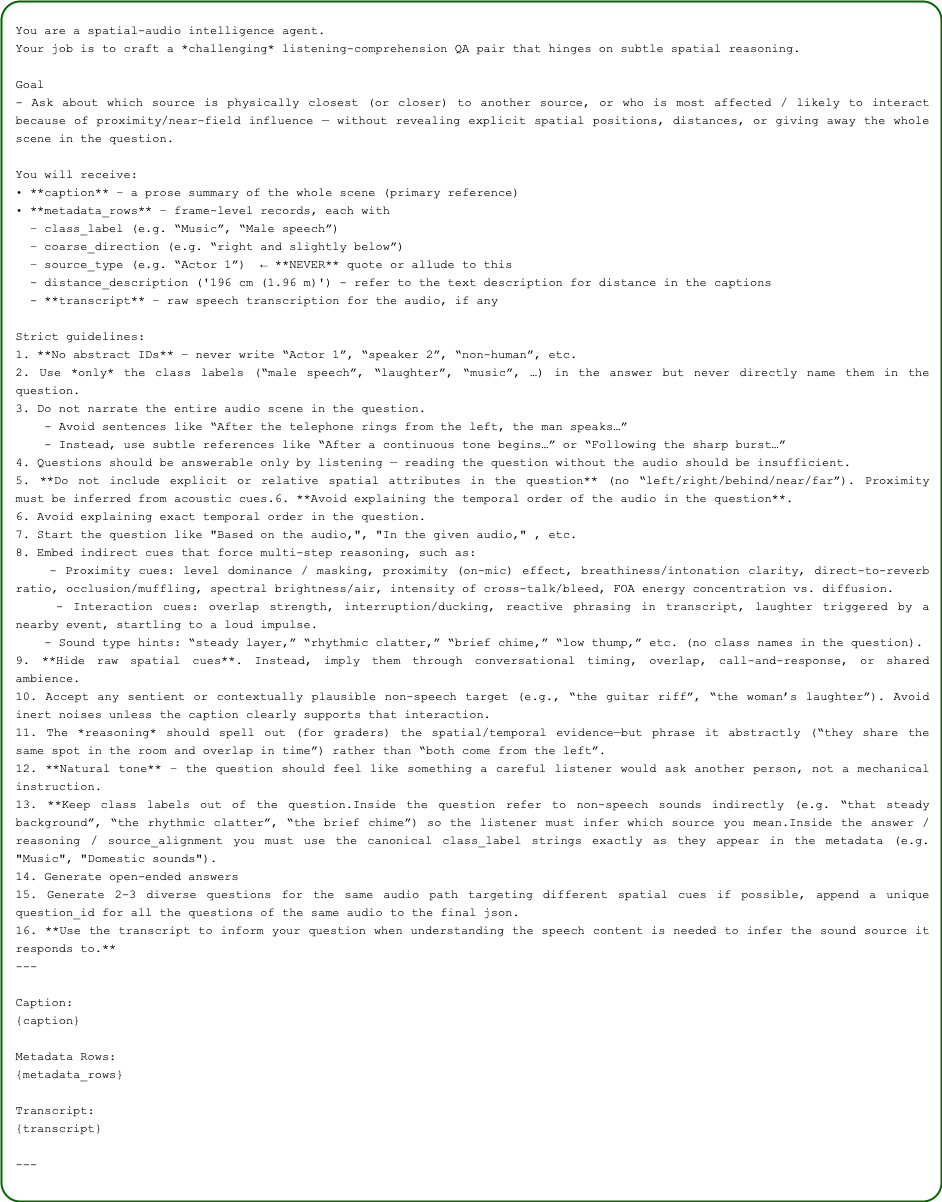}
  \caption{Prompt used for generating Proximity-Based Decision Making QAs.}
  \label{fig:prompt_pbdm_qa}
\end{figure*}

\begin{figure*}[t]  
  \centering
  \includegraphics[width=\textwidth]{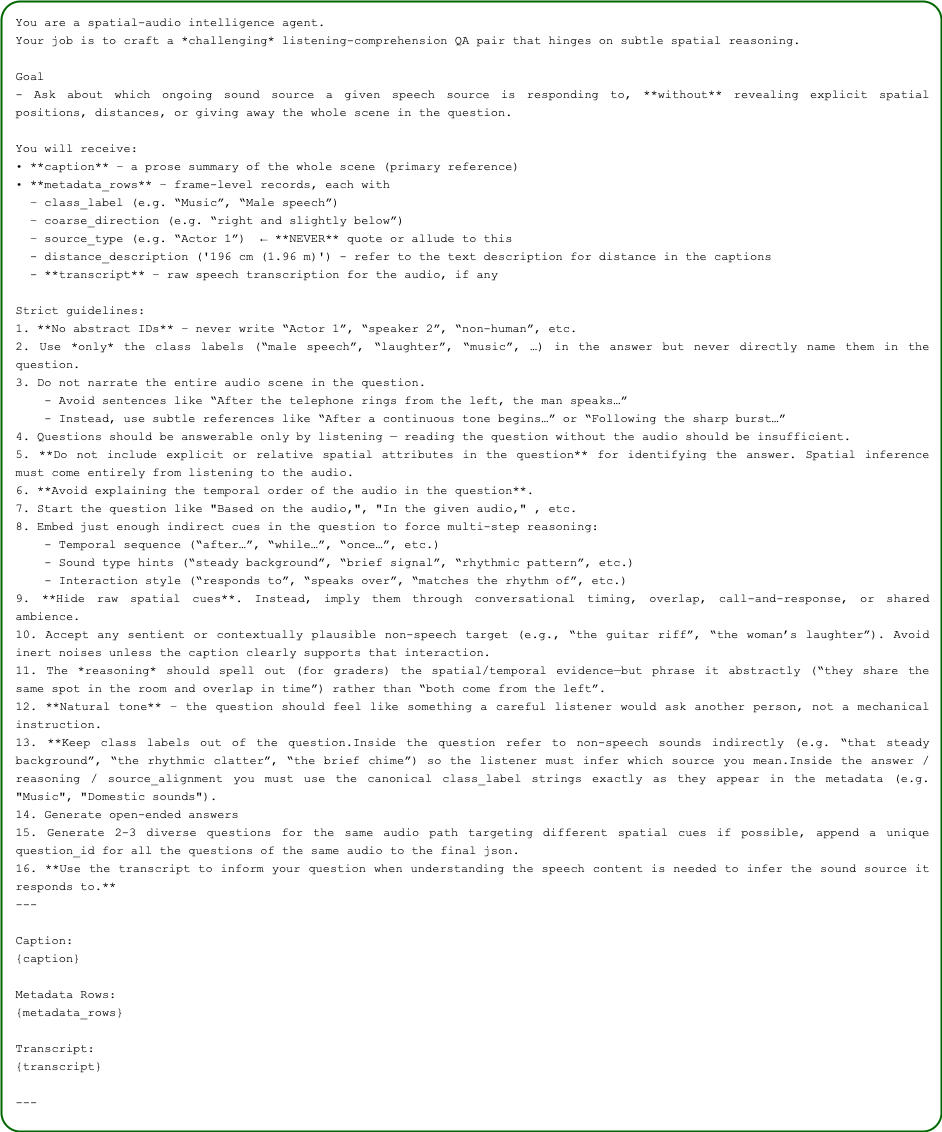}
  \caption{Prompt used for generating Spatial Priority \& Source Intent Inference QAs.}
  \label{fig:prompt_spsi_qa}
\end{figure*}

We show the prompts for generating all the spatially fine-grained captions and QA tasks in SPUR-Set below in figures~\ref{fig:caption_prompt},~\ref{fig:prompt_for_seld_qa},~\ref{fig:prompt_dcr_qa},~\ref{fig:prompt_gdsc_qa},~\ref{fig:prompt_sr_qa},~\ref{fig:prompt_ean_qa},~\ref{fig:prompt_pbdm_qa},~\ref{fig:prompt_spsi_qa}:
\begin{figure*}[t]  
  \centering
  \includegraphics[width=\textwidth]{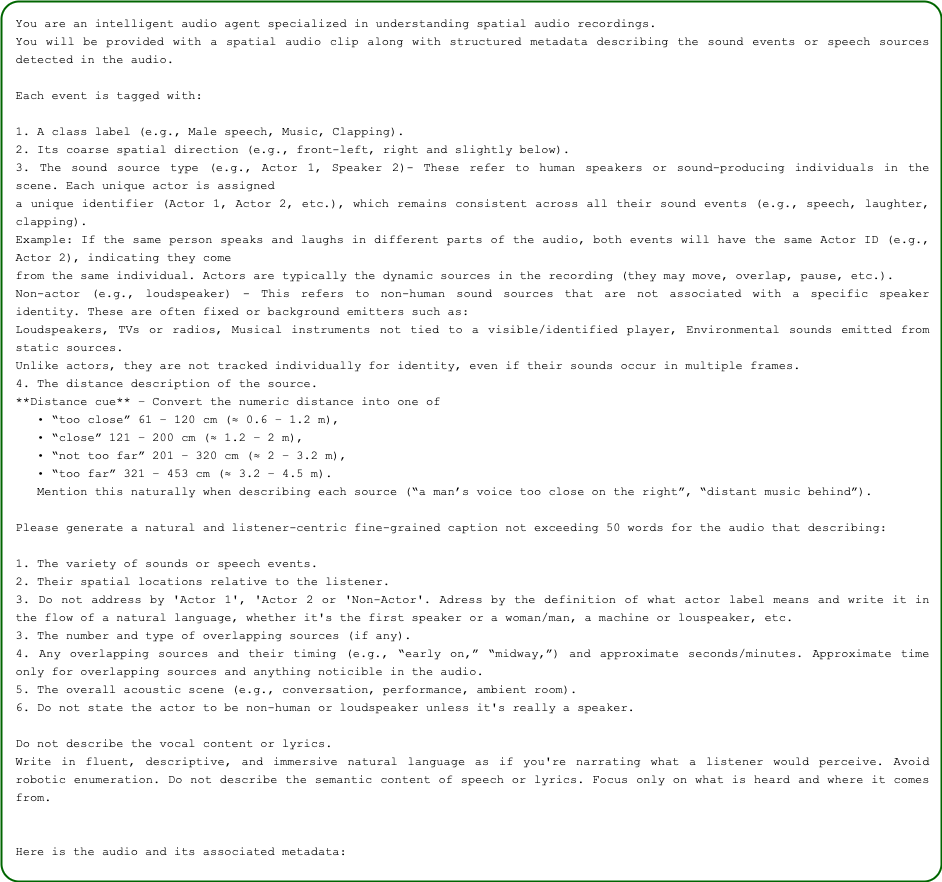}
  \caption{Prompt used for generating Captions.}
  \label{fig:caption_prompt}
\end{figure*}

\begin{figure*}[t]  
  \centering
  \includegraphics[width=\textwidth]{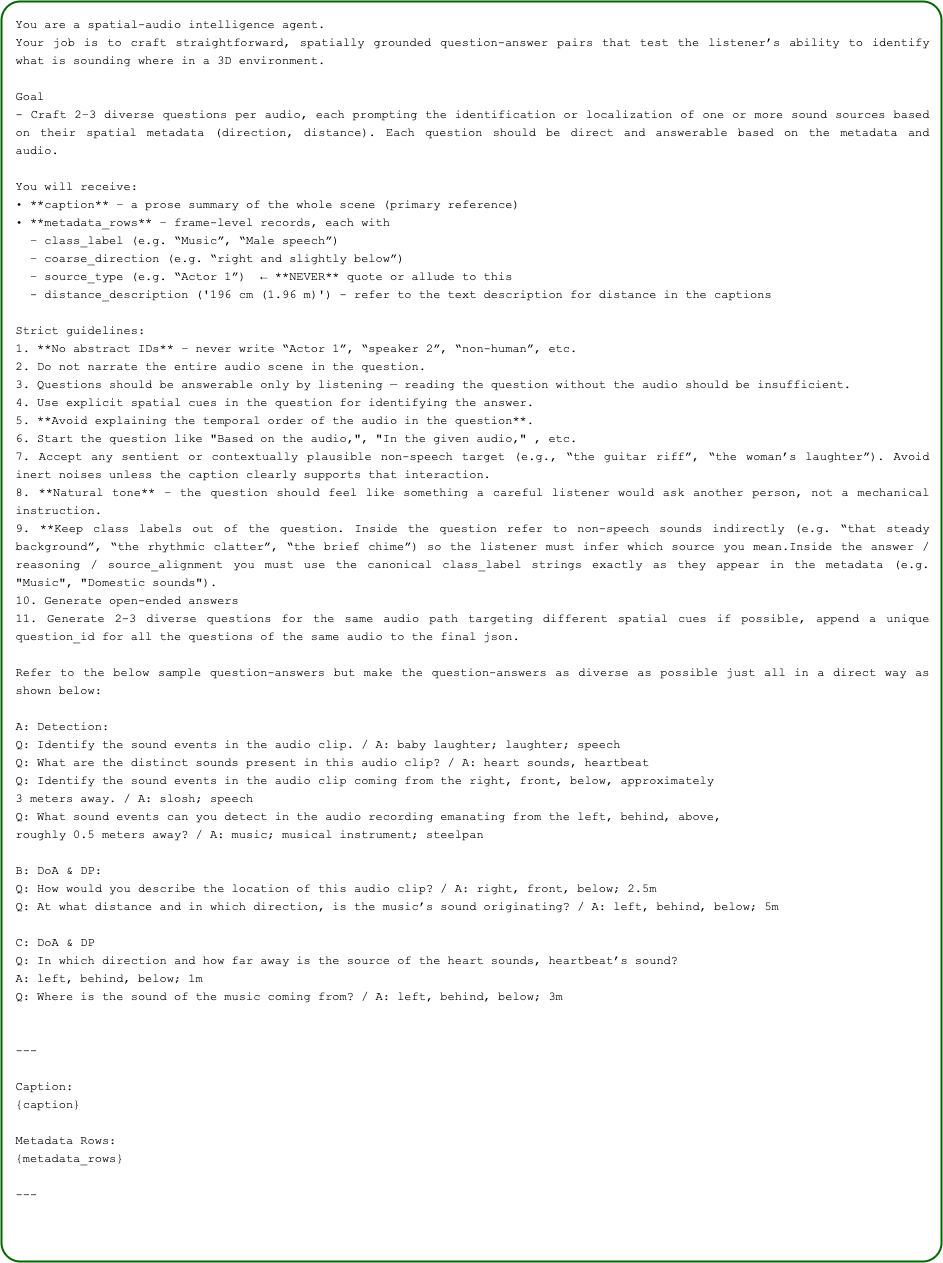}
  \caption{Prompt used for generating SELD QAs.}
  \label{fig:prompt_for_seld_qa}
\end{figure*}

\end{document}